\documentclass[seceq]{ptptex}

\notypesetlogo
\def\mbf{\boldsymbol}

\title{
General Relativistic Effects of Gravity in Quantum Mechanics
}
\subtitle{A Case of Ultra-Relativistic, Spin $1/2$ Particles }   

\author{
Kohkichi {\sc Konno}\footnote{Present address: 
Department of Physics, Hiroshima University, 
Higashi-Hiroshima 739-8526, Japan}
and Masumi {\sc Kasai}
}

\inst{
Faculty of Science and Technology, Hirosaki University, 
Hirosaki 036-8561, Japan
}

\recdate{
June 9, 1998
}

\abst{ We present a general relativistic framework for studying
gravitational effects in quantum mechanical phenomena. We concentrate
our attention on the case of ultra-relativistic, spin-$1/2$ particles
propagating in Kerr spacetime.  The two-component Weyl equation with
general relativistic corrections is obtained in the case of a slowly
rotating, weak gravitational field. Our approach is also applied to neutrino
oscillations in the presence of a gravitational field.  The relative
phase of two different mass eigenstates is calculated in radial
propagation, and the result is compared with those of the previous works. }

\begin{document}

\maketitle

\section{Introduction}

It had been thought that physical phenomena in which gravitational
effects and quantum effects appear simultaneously are far beyond 
our reach, before Colella, Overhauser and Werner \cite{cow} conducted 
an elegant experiment using a neutron interferometer.
(This kind of experiment is called a COW experiment.)
The COW experiment was the first experiment that measures 
the Newtonian gravitational effect on a wave function. 
This effect and its detectability were first suggested by
Overhauser and Colella,\cite{oc} and the next year 
the effect was verified by Colella et al.\cite{cow}
Although their analysis, which was based on inserting 
the Newtonian gravitational potential into the 
Schr\"odinger equation, was very simple, this experiment is 
conceptually very important in the history of quantum theory.

Recently, gravitational effects on another physical phenomenon, 
neutrino oscillations, have been much discussed. 
\cite{ab1,bhm,kojima,gl,cf,fgks} The COW experiment and this phenomenon
have the common aspect that gravitational effects appear in the quantum 
interference. However, there are some differences between
the two. In the former case, the spatial spread of the wave 
function plays a significant role, whereas in the latter
case, the existence of different mass eigenstates and linear
superposition are important. Another important difference is 
that the particle is non-relativistic in the former case, 
whereas it is ultra-relativistic in the latter case. 

It seems that a controversy concerning gravitationally induced neutrino
oscillation phases has arisen. Ahluwalia and Burgard \cite{ab1} state
that the phases amount to roughly 20\% of the kinematic
counterparts in the vicinity of a neutron star. Nevertheless,
the definition of neutrino energy and the derivation of the phases
are not clear in their original paper. \cite{ab1} On the other 
hand, other groups \cite{bhm,kojima,cf,fgks} have obtained
similar results for radially propagating neutrinos (the results 
seem to be different from that in Ref.~\citen{ab1}).  However,
the authors of Ref.~\citen{bhm} assume that different mass 
eigenstates are produced at different times. This assumption 
seems to be of questionable validity because the relative phase between the 
two different mass eigenstates initially becomes arbitrary. 
All of these papers, except Ref.~\citen{cf},
are based on a previous work, \cite{std}
in which the classical action is taken as a quantum phase.
Therefore, effects arising from the spin of the particle are not 
considered in these papers. On the other hand, the authors of 
Ref.~\citen{cf} use the covariant Dirac equation, but they also calculate 
the classical action along the particle trajectory in the end.

We provide another framework, different
from those of the previous works, for studying general 
relativistic gravitational effects on spin-1/2  
particles with non-vanishing mass, such as massive neutrinos.
(The experimental confirmation demonstrating that neutrinos have 
nonzero mass has not yet been obtained. However, recent experimental
reports \cite{totsuka} seem to suggest neutrinos to be massive.)
We do not merely calculate the classical action along the 
particle trajectory, but start from the covariant Dirac equation.
Our approach allows us to discuss the effects of the coupling
between the spin and the gravitational field.
In particular, we consider the propagation of the particle in the Kerr
geometry, by which the external field of a rotating star can be 
described. We perform our calculations in a slowly rotating, 
weak gravitational field, and derive the two-component Weyl equations
with corrections arising from the non-vanishing mass and the 
gravitational field. Furthermore, we discuss neutrino oscillations 
in the presence of the gravitational field.

The organization of this paper is as follows. 
In \S \ref{covariant-D-kerr}, we assume that the external
field of a rotating object is described by the Kerr metric
and discuss the covariant Dirac equation in this field.
In \S \ref{u-limit}, we derive the Weyl equations with
general relativistic corrections for an ultra-relativistic 
particle. The application to neutrino oscillations in the 
presence of the gravitational field is discussed  
in \S \ref{application}. Finally, we give a summary 
and conclusion in \S \ref{s-c}.

\section{Covariant Dirac equation in Kerr geometry}
\label{covariant-D-kerr}

In this section, we consider the covariant Dirac equation in the 
presence of a gravitational field arising from a rotating object.
We derive an equation for the time evolution of spinors, which 
describe particles with spin-1/2, in the last part of this section.

\subsection{Cavariant Dirac equation}

To begin, we briefly review the covariant Dirac equation. 
\cite{bw,weinberg2,bd}

The natural generalization of the Dirac equation into 
curved space-time gives
\begin{equation}
\label{covariant-dirac-eq}
 \left[ i \hbar \gamma^{\mu} \left( \frac{\partial}
     {\partial x^{\mu}} - \Gamma_{\mu} \right) - mc \right]
 \Psi = 0 ,
\end{equation}
where $\gamma^{\mu}$ are the covariant Dirac matrices connected with 
space-time through the relations
\begin{equation}
 \gamma^{\mu} \gamma^{\nu} + \gamma^{\nu} \gamma^{\mu}
   = 2 g^{\mu \nu} ,
\end{equation}
and $\Gamma_{\mu}$ is the spin connection. The spin connection
is determined by the condition
\begin{equation}
 \frac{\partial \gamma_{\nu}}{\partial x^{\mu}}
   - \Gamma^{\lambda}_{\ \nu \mu} \gamma_{\lambda}
   - \Gamma_{\mu} \gamma_{\nu} + \gamma_{\nu} \Gamma_{\mu} = 0 .
\end{equation}
We now introduce the constant Dirac matrices $\gamma^{(a)}$ 
defined by
\begin{equation}
 \gamma^{(a)} = e^{(a)}_{\ \ \mu} \gamma^{\mu} ,
\end{equation}
where $e^{(a)}_{\ \ \mu}$ is the orthogonal tetrad satisfying
the relation
\begin{equation}
 g_{\mu \nu} = \eta_{ab} e^{(a)}_{\ \ \mu} e^{(b)}_{\ \ \nu}
\end{equation}
($\eta_{ab} = \mbox{diag} \left( c^2 , -1 , -1 ,-1 \right)$).
Using these constant Dirac matrices, the spin connection is
expressed as
\begin{equation}
\label{spin-connection}
 \Gamma_{\mu} = - \: \frac{1}{8} \left[ \gamma^{(a)} , 
    \gamma^{(b)} \right] g_{\nu \lambda} \: e^{\ \ \nu}_{(a)}
    \nabla_{\mu} e^{\ \ \lambda}_{(b)} ,
\end{equation}
where the square brackets denote the usual commutator.\footnote{
We have ignored a term proportional to the unit matrix.
}

\subsection{Space-time}

Here, we discuss the gravitational field arising from a
rotating object. We now assume that the external field of 
the rotating object is described by the Kerr metric. 
If we restrict ourselves to a slowly rotating, weak gravitational
field up to the first order in the angular velocity, which 
is related to the Kerr parameter $a$, and the Newtonian 
gravitational potential $\phi = - GM / r$,
respectively, the line element is given by
\begin{eqnarray}
\label{metric}
  ds^2 & \simeq & \left( 1 + 2 \: \frac{\phi}{c^2} \right) c^2 
     dt^2 + \frac{4GMa}{c^2 r^3}
     \left( x dy - y dx \right) dt \nonumber \\
  & & \quad - \: \left( 1 - 2 \: \frac{\phi}{c^2} \right) 
     \left( dx^2 + dy^2 + dz^2 \right) ,
\end{eqnarray}
where $a$ is expressed in terms of the mass $M$ and 
the angular momentum $J$ of the gravitational source:
\begin{equation}
 a \equiv \frac{J}{M} .
\end{equation}
Assuming that the rotating object is a sphere of radius $R$ with
uniform density, we have 
\begin{equation}
\label{sphere-assumption}
  a \equiv \frac{J}{M} = \frac{2}{5} \: R^2 \omega ,
\end{equation}
where $\omega$ denotes the angular velocity of this object.
(If the rotating object deviates from a sphere, or has an 
inhomogeneous density distribution, then the numerical factor
$2/5$ might be changed by a factor of order unity.)

\subsection{Equation for time evolution of spinors}

Next, we turn our attention to time evolution of spinors.
The covariant Dirac equation (\ref{covariant-dirac-eq}) has beautiful
space-time symmetry. However, in order to investigate the time 
evolution of spinors, we must break the symmetrical form 
of this equation. For this purpose, we now use the (3+1) formalism.
In the (3+1) formalism, the metric $g_{\alpha \beta}$ is 
split as
\begin{subequations}
\begin{eqnarray}
  g_{00} & = & N^2 - \gamma_{ij} N^i N^j ,  \\
  g_{0i} & = & - \gamma_{ij} N^j \equiv - N_{i} , \\
  g_{ij} & = & - \gamma_{ij} ,
\end{eqnarray}
\end{subequations}
where $N$ is the lapse function, $N^i$ the shift vector,
and $\gamma_{ij}$ the spatial metric on the 3D hypersurface.
Furthermore, we define $\gamma^{ij}$ as the inverse matrix of 
$\gamma_{ij}$. Using the metric (\ref{metric}) derived in 
the last section, we can write the lapse function, the shift 
vector and the spatial metric in the following way:
\begin{eqnarray}
\label{lapse}
  N & = & c \left( 1 + \frac{\phi}{c^2} \right), \; \; 
\end{eqnarray}
\begin{subequations}
\begin{eqnarray}
  N^x & = & \frac{4GMR^2}{5c^2 r^3} \omega y ,    \\
  N^y & = & - \frac{4GMR^2}{5c^2 r^3} \omega x ,    \\
  N^z & = & 0 , 
\end{eqnarray}
\end{subequations}
\begin{eqnarray}
\label{smetric}
 \: \gamma_{ij} & = & \left( 1 - 2 \frac{\phi}{c^2} \right) \delta_{ij}.
\end{eqnarray}

Furthermore, we choose the tetrad as
\begin{eqnarray}
  e^{\ \ \mu}_{(0)} & = & c \left( \frac{1}{N}, - \frac{N^i}{N} \right), \\
  e^{\ \ \mu}_{(k)} & = & \left( 0 , e^{\ \ i}_{(k)} \right) ,
\end{eqnarray}
where the spatial triad $e^{\ \ i}_{(k)}$ is defined as
\begin{equation}
  \gamma_{ij} \: e^{\ \ i}_{(k)} \: e^{\ \ j}_{(l)} = \delta_{kl} .
\end{equation}
       From Eqs. (\ref{lapse})--(\ref{smetric}), we derive
\begin{subequations}
\begin{eqnarray}
\label{cs-of-tetrad-s}
  e_{(0)}^{\ \ 0} & = & 1 - \frac{\phi}{c^2} , \\
  e_{(0)}^{\ \ 1} & = & 
     - \frac{4GMR^2}{5 c^2 r^3} \omega y , \\
  e_{(0)}^{\ \ 2} & = & 
     \frac{4GMR^2}{5 c^2 r^3} \omega x , \\
  e_{(0)}^{\ \ 3} & = & 0 , 
\end{eqnarray}
\end{subequations}
\begin{eqnarray}
\label{cs-of-tetrad-g}
  e_{(j)}^{\ \ i} & = & \left( 1 + \frac{\phi}{c^2} \right) 
     \delta_{j}^{\ i}. \quad
\end{eqnarray}
Using our choice of the tetrad, the covariant Dirac matrices 
$\gamma^{\alpha}$ are written as
\begin{eqnarray}
  \gamma^0 & = & \gamma^{(a)} e^{\ \ 0}_{(a)}
     \; = \; \gamma^{(0)} \frac{c}{N} ,\\
  \gamma^i & = & \gamma^{(a)} e^{\ \ i}_{(a)}
     \; = \; - \: \gamma^{(0)} \frac{c}{N} N^i 
     + \gamma^{(j)} e^{\ \ i}_{(j)} .
\end{eqnarray}
Hence the covariant Dirac equation (\ref{covariant-dirac-eq}) 
can be written as
\begin{eqnarray}
\label{h-general-form}
  \! \! i \hbar \frac{\partial}{\partial t} \Psi
  & = & H \Psi \nonumber \\
  & = & \left[ \left( \gamma^{(0)} \gamma^{(j)} c N e^{\ \ i}_{(j)}
     - N^i \right) \left( \overline{p}_{i} + i \hbar \Gamma_{i}
     \right) + i \hbar \Gamma_{0} + \gamma^{(0)} mc^2 N \right] \Psi ,
\end{eqnarray}
where $\overline{p}_{i}$ is the momentum operator in flat space-time.
This equation describes the time evolution of spinors. 
If we adopt the Weyl representation as the constant Dirac
matrices in this equation, then for massless particles 
in flat space-time we derive the well-known Weyl equations
\begin{equation}
  i \hbar \frac{\partial}{\partial t} \psi
  = \pm c {\mbf \sigma} \cdot \overline{\mbf p} \psi ,
\end{equation}
where $\psi$ denotes two-component spinors.

\section{Ultra-relativistic limit}
\label{u-limit}

We now restrict our attention to the ultra-relativistic limit, which
means that the rest energy of the particle is much smaller than 
the kinetic energy in the observer's frame. In particular, we
expand the energy of the particle itself up to 
$O \left( m^2 c^4 / pc \right)$.

Here, we obtain the ultra-relativistic Hamiltonian up to 
the order of interest by performing a unitary 
transformation similar to the Foldy-Wouthuysen-Tani (FWT) 
transformation.\cite{fw,tani}

First, following the discussion of Ref.~\citen{wkf}, we 
redefine the spinor and the Hamiltonian according to
\begin{equation}
\label{redefinition}
  \Psi' = \gamma^{1/4} \Psi , \quad 
  H' = \gamma^{1/4} H \gamma^{-1/4} ,
\end{equation}
where $\gamma$ is the determinant of the spatial metric:
\begin{equation}
 \gamma = \det \left( \gamma_{ij} \right) .
\end{equation}
Since the invariant scalar product is
\begin{equation}
 \left( \psi , \varphi \right)
   \equiv \int \! \overline{\psi} \varphi 
   \sqrt{\gamma} d^3 x ,
\end{equation}
under this redefinition the scalar product comes to assume the 
same form as in flat space-time:
\begin{equation}
 \left< \psi' , \varphi' \right> 
    \equiv \int \! \overline{\psi'} \varphi' d^3 x .
\end{equation}
It is sometimes convenient to adopt this definition of 
the scalar product.

Next, we perform a unitary transformation to derive the 
ultra-relativistic Hamiltonian which is the ``even'' operator
up to the order of our interest. From this, we have 
\begin{eqnarray}
  \tilde{H'}
  & = & U H' U^{\dag} \nonumber \\
  & = & \left( 
      \begin{array}{cc}
        H_{R} & 0 \\ 0 & H_{L}
      \end{array}
      \right) + \left[ O \left( \frac{m^3 c^6}{p^2 c^2} \right) 
      \; \mbox{or} \; \: O \left( \frac{\phi^2}{c^4}, 
      \omega^2 \right) \right] .
\end{eqnarray}
The Dirac spinor is also divided into each of two-component spinors as
\begin{equation}
  \tilde{\Psi'} = \left( 
    \begin{array}{c}
      \psi_{R} \\  \psi_{L}
    \end{array}
    \right) ,
\end{equation}
where the subscript $R$ and $L$ denote the right-handed and the left-handed
components, respectively.

We consider the left-handed component. 
We find that the equation for this 
component is given by
\begin{eqnarray}
\label{ultra-rel-h1}
  i \hbar \frac{\partial}{\partial t} \psi_{L}
  & = & H_{L} \psi_L  \nonumber \\
  & = &  - \left[ \left\{ 1 + \frac{1}{c^2} \left( \phi
    + \overline{\mbf p} \cdot \phi \overline{\mbf p}
    \frac{1}{\overline{p}^2} + 2 \frac{GM}{r^3} {\mbf L}
    \cdot {\mbf S} \frac{1}{\overline{p}^2} 
    \right) \right\} c \overline{p} \;
    \frac{{\mbf \sigma} \cdot \overline{\mbf p}}{\overline{p}}
    \right. \nonumber \\
  & & \quad - \: \frac{1}{c^2} \left( \frac{4GMR^2}{5r^3} 
    {\mbf \omega} \cdot \left( {\mbf L} + {\mbf S} \right)
    + \frac{6GMR^2}{5r^5} {\mbf S} \cdot \left[ {\mbf r} \times
    \left( {\mbf r} \times {\mbf \omega} \right) \right]
    \right) \nonumber \\
  & & \quad + \: \left\{ 
    1 + \frac{1}{4 c^2} \left( \phi - \frac{1}{\overline{p}^2}
    \phi \overline{p}^2 + \frac{1}{\overline{p}^2} \overline{\mbf p}
    \cdot \phi \overline{\mbf p} - \overline{\mbf p}
    \cdot \phi \overline{\mbf p} \frac{1}{\overline{p}^2}
    \right. \right. \nonumber \\
  & & \qquad \qquad \qquad \quad \left. \left. 
    + \: 2 \frac{1}{\overline{p}^2} \frac{GM}{r^3} {\mbf L} \cdot
    {\mbf S} - 2 \frac{GM}{r^3} {\mbf L} \cdot {\mbf S}
    \frac{1}{\overline{p}^2}
    \right) \right\} \frac{m^2 c^3}{2 \overline{p}} \;
    \frac{{\mbf \sigma} \cdot \overline{\mbf p}}{\overline{p}}
    \nonumber \\
  & & \left. \quad + \: \frac{1}{8} m^2 c^2 \: \frac{1}{c^2} \left( 
    A \frac{1}{\overline{p}^2} - 2 \frac{{\mbf \sigma} \cdot 
    \overline{\mbf p}}{\overline{p}^2} A
    \frac{{\mbf \sigma} \cdot \overline{\mbf p}}
    {\overline{p}^2} + \frac{1}{\overline{p}^2} A
    \right) \right] \psi_{L} , 
\end{eqnarray}
where
\begin{equation}
\label{a-definition}
  A = \frac{4GMR^2}{5r^3} 
    {\mbf \omega} \cdot \left( {\mbf L} + {\mbf S} \right)
    + \frac{6GMR^2}{5r^5} {\mbf S} \cdot \left[ {\mbf r} \times
    \left( {\mbf r} \times {\mbf \omega} \right) \right] .
\end{equation}
The details of the calculations are given in Appendix
\ref{apd:d-ultra-H}. From this, we find how the spin-orbit 
coupling, the coupling between the spin and the rotation
of the gravitational source, or the coupling between the 
total angular momentum and the rotation
is coupled to the non-vanishing mass. 

In radial propagation, the orbital angular momentum vanishes. 
Therefore, in this case, only spin effects coupled to the 
rotation appear. If we set ${\mbf \omega} = {\bf 0}$, then 
there is no spin effect in radial propagation. This consequence 
is consistent with the results of previous work.\cite{bw}

\section{An application}
\label{application}

In this section, we consider an application of the two-component
equation derived in the last section to neutrino oscillations in
the presence of a gravitational field. In neutrino oscillations
(see, e.g., Refs.~\citen{kayser,bp,bahcall} for analysis
in flat space-time), the most important point is the phase difference
of the two different mass eigenstates. Hence we now concentrate
on the phase shift of the particle. 

\subsection{Neutrino oscillation in Kerr space-time}

We derive the phase shift directly from the 
two-component equation derived in the last section.
Furthermore, for simplicity, we consider the radial 
propagation ($r$-direction), 
in which the spin-orbit coupling vanishes.

We now regard terms arising from the non-vanishing mass and the gravitational
field as perturbations. Then Eq.~(\ref{ultra-rel-h1})
for the left-handed component is considered to be
\begin{equation}
\label{perturbed-eq}
 i \hbar \frac{\partial}{\partial t} \psi_{L}
   = \left( H_{0L} + \Delta H_{L} \right) \psi_{L} ,
\end{equation}
where $H_{0L}$ denotes the unperturbed Hamiltonian 
$H_{0L} = - c {\mbf \sigma} \cdot \overline{\mbf p}$,
and $\Delta H_{L}$ the corrections arising from the non-vanishing 
mass and the gravitational field.

Here we assume that the spinor $\psi_{L}$ is given by
\begin{equation}
\label{psi-psi0}
 \psi_{L} \left( {\mbf x} , t \right)
   = e^{i \Phi (t)} \psi_{0L} \left( {\mbf x} , t \right) ,
\end{equation}
where $\psi_{0L}$ satisfies the equation
\begin{equation}
\label{psi0-eq}
 i \hbar \frac{\partial}{\partial t} 
    \psi_{0L} \left( {\mbf x} , t \right)
 = H_{0L} \psi_{0L} \left( {\mbf x} , t \right) 
\end{equation}
Substituting Eq.~(\ref{psi-psi0}) into Eq.~(\ref{perturbed-eq})
and using Eq.~(\ref{psi0-eq}), we obtain 
\begin{equation}
 \Phi = - \frac{1}{\hbar} \int^{t} \! \Delta H_{L} dt .
\end{equation}

In order to derive the gravitationally induced phases practically,
we assume that corresponding to the left-handed component, 
$\psi_{0L}$ satisfies the relation
\begin{equation}
 \frac{{\mbf \sigma} \cdot \overline{\mbf p}}{\overline{p}}
   \psi_{0L} \left( {\mbf x} , t \right)
 = - \: \psi_{0L} \left( {\mbf x} , t \right) .
\end{equation}
Furthermore, we here replace the {\it q}-numbers in $\Delta H_{L}$
with {\it c}-numbers. This is a kind of semi-classical 
approximation.  From this, excluding spin effects, we derive the phase
\begin{equation}
 \Phi = - \frac{1}{\hbar} \int_{t_{A}}^{t_{B}}
   \left( 2 \frac{\phi}{c^2} cp + \frac{m^2 c^3}{2p} \right) dt,
\end{equation}
where we have considered the case that the neutrino is produced
at a space-time point $A (t_{A},r_{A})$ and detected at a 
space-time point $B (t_{B},r_{B})$. We now concentrate on
the term related to $m^2$, because neutrino oscillations take 
place as a result of the mass square difference. Let the two different
mass eigenstates have common momentum $p$ and propagate along the
same path. Then the relative phase $\Delta \Phi_{ij}$ of the two
different mass eigenstates, $\left| \nu_{i} \right>$ and 
$\left| \nu_{j} \right>$, is given by
\begin{equation}
\label{result-no-spin}
 \Delta \Phi_{ij} = \frac{\Delta m^2_{ij} c^3}{2 \hbar}
    \int_{t_{A}}^{t_{B}} \! \frac{1}{p} dt .
\end{equation}

Next, we discuss the spin-rotation coupling.
In a similar way, we replace the {\it q}-numbers in terms
concerning the spin of the particle in Eq.~(\ref{ultra-rel-h1})
with {\it c}-numbers again. From this kind of semi-classical 
approximation, we find that the spin-rotation term coupled to 
$m^2$ vanishes. However, the spin-rotation effects in higher 
order terms may survive. This fact implies that there is no influence 
of the spin-rotation coupling on neutrino oscillations, at least
up to the order related to $m^2$.

Consequently, in a radially propagating case, we obtain 
the phase difference (\ref{result-no-spin})
between the two different mass eigenstates as a final result.

\subsection{Comparison with previous works}

Finally, we compare the result obtained above with those of previous works. 
For this purpose, we consider the Schwarzschild limit (i.e.,  
${\mbf \omega} \rightarrow 0$), which leads to the result 
(\ref{result-no-spin}) again.

First, we assume the ``background'' neutrino trajectory as that
of radial null geodesics:
\begin{equation}
 0 = ds^2 = \left( 1 + 2 \frac{\phi}{c^2} \right) 
   c^2 dt^2 - \left( 1 - 2 \frac{\phi}{c^2} \right) dr^2 .
\end{equation}
Then we obtain 
\begin{equation}
 dt \simeq \frac{1}{c} \left( 1 - 2 \frac{\phi}{c^2} \right) dr .
\end{equation}
Hence, if we transform the integral (\ref{result-no-spin})
with respect to $t$ to that with respect to $r$, 
the relative phase $\Delta \Phi_{ij}$ is given by
\begin{equation}
 \Delta \Phi_{ij} = \frac{\Delta m^2_{ij} c^3}{2 \hbar}
   \int_{r_A}^{r_B} \frac{1}{pc} \left( 1 - 2 \frac{\phi}{c^2} \right) dr .
\end{equation}
The second term in the round brackets corresponds to the gravitational 
correction as indicated by Ahluwalia and Burgard.\cite{ab1} Indeed, 
under the assumption that the tetrad component of the radial momentum
$p_{(r)} = e_{(r)}^{\ \ r} p_{r}$ is constant along the trajectory,
we can obtain the same expression as in Ref.~\citen{ab1}. 
(Note that Ahluwalia and Burgard\cite{ab1} assume $p_{(r)} c$ as the energy
of the neutrino.)

Next, let us see whether our result (\ref{result-no-spin}) reproduces
the other form of the results.\cite{kojima,cf,fgks}  From the mass
shell condition  
$g^{\mu \nu} p_{\mu} p_{\nu} = m^2 c^2$, 
$p$ is related with the energy $E ( \equiv p_{t} c )$ in
the following way: 
\begin{equation}
p c = \left( 1 - 2 \frac{\phi}{c^2} \right) E + \left[ O\left(m^2\right) 
  \ \mbox{or}\  O\left(\phi^2\right)\right]. 
\end{equation}
Under the assumption that $E$ is constant along the trajectory, we
finally obtain 
\begin{eqnarray}
 \Delta \Phi_{ij} & = & \frac{\Delta m^2_{ij} c^3}{2 \hbar}
    \int_{r_{A}}^{r_{B}} \! \frac{dr}{E} \nonumber \\
  & = & \frac{\Delta m^2_{ij} c^3}{2 \hbar E} 
    \left( r_{B} - r_{A}\right) , 
\end{eqnarray}
which is the same as the result obtained in previous 
works.\cite{kojima,cf,fgks}
In this sense, our result Eq.~(\ref{result-no-spin}) contains both of
the previous expressions. Our analysis here clearly shows that 
the controversy arising due to the discrepancy
between the results of Ahluwalia and
Burgard\cite{ab1} and other authors\cite{kojima,cf,fgks} is simply
due to different assumptions concerning constancy along the neutrino
trajectory.

\section{Summary and conclusion}
\label{s-c}

We have studied the general relativistic effects of gravity on
spin-1/2 particles with non-vanishing mass. In particular, we have
considered particles propagating in the Kerr geometry in the
slowly rotating, weak field approximation. By performing a unitary
transformation similar to the FWT transformation, we have obtained the
two-component Weyl equations with the corrections arising from the
non-vanishing mass and the gravitational field from the covariant
Dirac equation. The Hamiltonian clearly shows how the spin-orbit
coupling, the spin-rotation coupling or the coupling between the total
angular momentum and the rotation is coupled to the non-vanishing
mass.

Furthermore, we have discussed an application of the two-component
equations to neutrino oscillations in the presence of a
gravitational field, and we have derived the phase difference of the two
different mass eigenstates in radial propagation.  It is worth
mentioning that our result contains both of the previous
expressions. We have shown that the controversy arising
from the discrepancy
between the results of Ahluwalia and Burgard\cite{ab1} and
other authors\cite{kojima,cf,fgks} is simply due to different
assumptions regarding constancy along the neutrino trajectory. Moreover, as
seen in the transformation of the integral variable, the
gravitationally induced neutrino oscillation phases arise from the
modification of the propagating distance. Indeed, we found that the
gravitational correction term comes out in this variable
transformation.

We have not applied our approach to a non-radially propagating case in
detail in this paper. However, it is of interest whether the
spin-orbit coupling affects the neutrino oscillations. This will be the
subject of further investigation.

Although it seems difficult to provide verification of these
effects with current experimental detectability, we believe that
investigation of systems in which both quantum effects and
gravitational effects come into play is important. Progress in
technology may make the verification of such effects possible.

\section*{Acknowledgements}

We would like to thank H.~Asada and T.~Futamase for useful suggestions
and discussions and Y.~Kojima for valuable comments.

\appendix
\section{Derivation of Ultra-Relativistic Hamiltonian}
\label{apd:d-ultra-H}

\subsection{Components of spin connection}

The spin connection is given by Eq.~(\ref{spin-connection}):
\begin{equation}
 \Gamma_{\mu} = - \: \frac{1}{8} \left[ \gamma^{(a)} ,
   \gamma^{(b)} \right] g_{\lambda \sigma} \: e_{(a)}^{\ \ \lambda}
   \nabla_{\mu} e_{(b)}^{\ \ \sigma}.
\end{equation}
It is convenient to introduce the following $4 \times 4$ matrices
similar to the Pauli spin matrices:
\begin{equation}
 \rho_{1} = \left( 
   \begin{array}{cc}
     0 & I \\ I & 0 
   \end{array}
   \right), \quad 
 \rho_{2} = \left( 
   \begin{array}{cc}
     0 & -iI \\ iI & 0 
   \end{array}
   \right), \quad
 \rho_{3} = \left( 
   \begin{array}{cc}
     I & 0 \\ 0 & -I 
   \end{array}
   \right),
\end{equation}
where $I$ is the $2 \times 2$ unit matrix. These matrices satisfy the
relations
\begin{equation}
\label{rho-rel1}
 \rho_{i} \rho_{j} = \delta_{ij} + i \varepsilon_{ijk} \rho_{k} ,
\end{equation}
where $\varepsilon_{ijk}$ is the Levi-Civita antisymmetric tensor
($\varepsilon_{123} = +1$). If we adopt the Weyl representation as 
the constant Dirac matrices, then we have
\begin{equation}
 \gamma^{(0)} = \frac{1}{c} \rho_{1} , \quad 
 \gamma^{(i)} = - i \rho_{2} \sigma_{i} ,
\end{equation}
where $\sigma$ are the Pauli spin matrices.

Using the above quantities, the components of spin connection, 
up to the order of interest, are given by
\begin{eqnarray}
  i \hbar \Gamma_{0}
  & = & \frac{1}{2c} \: \rho_3 \: {\mbf \sigma} \cdot 
        \left( \overline{\mbf p} \phi \right) \nonumber \\
  & & \quad + \: \frac{1}{c^2} \left[ \frac{4GMR^2}{5r^3}
      {\mbf \omega} \cdot {\mbf S} + 
      \frac{6GMR^2}{5r^5} {\mbf S} \cdot \left[ {\mbf r}
      \times \left( {\mbf r} \times {\mbf \omega} \right)
      \right] \right] , \\
  i \hbar \Gamma_{1}
  & = & - \: \frac{\hbar}{2c^2} \left( \phi_{,2} \: \sigma_{3}
        - \phi_{,3} \: \sigma_2 \right) \nonumber \\
  & & \quad + \: \frac{i \hbar}{c^3} \: \rho_3 \frac{3GMR^2}{5r^5} 
      \omega \left[ - 2xy \: \sigma_{1} + \left( x^2 - y^2 \right) 
      \sigma_{2} - yz \: \sigma_{3} \right] , \\
  i \hbar \Gamma_{2}
  & = & - \: \frac{\hbar}{2c^2} \left( \phi_{,3} \: \sigma_{1}
        - \phi_{,1} \: \sigma_3 \right) \nonumber \\
  & & \quad + \: \frac{i \hbar}{c^3} \: \rho_3 \frac{3GMR^2}{5r^5} 
      \omega \left[ \left( x^2 - y^2 \right) \sigma_{1} 
      + 2 xy \: \sigma_{2} + zx \: \sigma_{3} \right] , \\
  i \hbar \Gamma_{3}
  & = & - \: \frac{\hbar}{2c^2} \left( \phi_{,1} \: \sigma_{2}
        - \phi_{,2} \: \sigma_1 \right) \nonumber \\
  & & \quad + \: \frac{i \hbar}{c^3} \: \rho_3 \frac{3GMR^2}{5r^5} 
      \omega \left[ - yz \: \sigma_{1} + zx \: \sigma_{2} \right] .
\end{eqnarray}

\subsection{Unitary transformation}

The Hamiltonian defined in Eq.~(\ref{h-general-form})
is given by
\begin{eqnarray}
  H & = & \rho_3 c {\mbf \sigma} \cdot \overline{\mbf p} 
    + \rho_3 \left[ - \frac{1}{2} c {\mbf \sigma} \cdot \left( 
    \overline{\mbf p} \frac{\phi}{c^2} \right) 
    + 2 \frac{\phi}{c^2} \: c {\mbf \sigma} \cdot \overline{\mbf p}
    \right] \nonumber \\
  & & \quad + \: \frac{1}{c^2} \left[ \frac{4GMR^2}{5r^3} 
    {\mbf \omega} \cdot \left( {\mbf L} + {\mbf S} \right)
    + \frac{6GMR^2}{5r^5} {\mbf S} \cdot \left[ {\mbf r}
    \times \left( {\mbf r} \times {\mbf \omega} \right) \right]
    \right] + \rho_1 mc^2 + \rho_1 mc^2 \frac{\phi}{c^2} .
    \nonumber \\
\end{eqnarray}
Moreover, the Hamiltonian redefined by Eq.~(\ref{redefinition}) is then
\begin{eqnarray}
  H' & = & \rho_3 c {\mbf \sigma} \cdot \overline{\mbf p} 
    + \rho_3 \left( c {\mbf \sigma} \cdot  
    \overline{\mbf p} \: \frac{\phi}{c^2}  
    + \frac{\phi}{c^2} \: c {\mbf \sigma} \cdot \overline{\mbf p}
    \right) \nonumber \\
  & & \: + \: \frac{1}{c^2} \left[ \frac{4GMR^2}{5r^3} 
    {\mbf \omega} \cdot \left( {\mbf L} + {\mbf S} \right)
    + \frac{6GMR^2}{5r^5} {\mbf S} \cdot \left[ {\mbf r}
    \times \left( {\mbf r} \times {\mbf \omega} \right) \right]
    \right] + \rho_1 mc^2 + \rho_1 mc^2 \frac{\phi}{c^2} ,
    \nonumber \\
\end{eqnarray}
where terms proportional to $\rho_{1}$ and $\rho_{2}$ are ``odd'',
and those proportional to $\rho_{3}$ are ``even''.

Next, by performing a unitary transformation similar to the FWT 
transformation, let us derive the ultra-relativistic Hamiltonian 
for the left-handed component. Here we divide the unitary 
transformation into several steps. First, we use the unitary operator
\begin{equation}
  U_1 = \exp \left( i \rho_2 \frac{1}{2} mc^2 
    \frac{c {\mbf \sigma} \cdot \overline{\mbf p}}
    {c^2 \overline{p}^2} \right) ,
\end{equation}
which is introduced to eliminate the odd term $\rho_{1} mc^2$.
Using the useful formula 
\begin{equation}
\label{formula}
  e^{iS} H e^{-iS}
  = H + i \left[ S , H \right] + \frac{i^2}{2!} \left[ S ,
    \left[ S , H \right] \right] + \frac{i^3}{3!} \left[ S ,
    \left[ S , \left[ S , H \right] \right] \right] + \cdots 
\end{equation}
and the relation (\ref{rho-rel1}),
we obtain the transformed Hamiltonian
\begin{eqnarray}
\label{f-h-ultra-rel}
  \lefteqn{U_{1} H' U^{\dag}_{1}} \nonumber \\
  & = & \rho_3 c {\mbf \sigma} \cdot \overline{\mbf p} 
    + \rho_3 \left( c {\mbf \sigma} \cdot  
    \overline{\mbf p} \: \frac{\phi}{c^2}  
    + \frac{\phi}{c^2} \: c {\mbf \sigma} \cdot \overline{\mbf p}
    \right) + \frac{1}{c^2} A + \rho_1 mc^2 \frac{\phi}{c^2} \nonumber \\
  & & - \: \rho_1 \frac{1}{2} mc^2 \left[ c {\mbf \sigma} \cdot 
    \overline{\mbf p} \left( \frac{c {\mbf \sigma}
    \cdot \overline{\mbf p}}{c^2 \overline{p}^2} \: \frac{\phi}{c^2}
    + \frac{\phi}{c^2} \: \frac{c {\mbf \sigma} \cdot \overline{\mbf p}}
    {c^2 \overline{p}^2} \right) 
    + \left( \frac{c {\mbf \sigma} \cdot \overline{\mbf p}}
    {c^2 \overline{p}^2} \: \frac{\phi}{c^2} + \frac{\phi}{c^2} \: 
    \frac{c {\mbf \sigma} \cdot \overline{\mbf p}}
    {c^2 \overline{p}^2} \right) c {\mbf \sigma} \cdot \overline{\mbf p} 
    \right] \nonumber \\
  & & + \: i \rho_2 \frac{1}{2} mc^2 \left( \frac{c {\mbf \sigma} 
    \cdot \overline{\mbf p}}{c^2 \overline{p}^2} \frac{A}{c^2}
    - \frac{A}{c^2} \frac{c {\mbf \sigma} \cdot \overline{\mbf p}}
    {c^2 \overline{p}^2} \right) + \rho_3 \frac{1}{2} m^2 c^4
    \frac{c {\mbf \sigma} \cdot \overline{\mbf p}}
    {c^2 \overline{p}^2} \nonumber \\
  & & - \: \rho_3 \frac{1}{8} m^2 c^4 \left[ \frac{1}
    {c^2 \overline{p}^2} \frac{\phi}{c^2} c {\mbf \sigma} \cdot
    \overline{\mbf p} + c {\mbf \sigma} \cdot \overline{\mbf p}
    \frac{\phi}{c^2} \frac{1}{c^2 \overline{p}^2} 
    - \left( \frac{c {\mbf \sigma} \cdot \overline{\mbf p}}
    {c^2 \overline{p}^2} \frac{\phi}{c^2} + \frac{\phi}{c^2}
    \frac{c {\mbf \sigma} \cdot \overline{\mbf p}}
    {c^2 \overline{p}^2} \right) \right] \nonumber \\
  & & - \: \frac{1}{8} m^2 c^4 \left( \frac{A}{c^2} 
    \frac{1}{c^2 \overline{p}^2} 
    - 2 \frac{c {\mbf \sigma} \cdot \overline{\mbf p}}
    {c^2 \overline{p}^2} \frac{A}{c^2} \frac{c {\mbf \sigma} 
    \cdot \overline{\mbf p}}
    {c^2 \overline{p}^2} + \frac{1}{c^2 \overline{p}^2} 
    \frac{A}{c^2} \right) ,
\end{eqnarray}
where $A$ is given by Eq.~(\ref{a-definition}):
\begin{equation}
  A = \frac{4GMR^2}{5r^3} {\mbf \omega} \cdot 
    \left( {\mbf L} + {\mbf S} \right) + \frac{6GMR^2}{5r^5} {\mbf S} 
    \cdot \left[ {\mbf r} \times \left( {\mbf r} \times {\mbf \omega} 
    \right) \right] .
\end{equation}
Second, in order to eliminate the second line in Eq. (\ref{f-h-ultra-rel}),
we use the unitary operator
\begin{equation}
  U_{2} = \exp \left[ - i \rho_2 \frac{1}{2} mc^2 \left(
    \frac{c {\mbf \sigma} \cdot \overline{\mbf p}}
    {c^2 \overline{p}^2} \frac{\phi}{c^2} + \frac{\phi}{c^2}
    \frac{c {\mbf \sigma} \cdot \overline{\mbf p}}
    {c^2 \overline{p}^2} \right) \right] .
\end{equation}
Using this unitary operator, we obtain
\begin{eqnarray}
  \lefteqn{U_{2} U_{1} H' U^{\dag}_{1} U^{\dag}_{2}}
    \nonumber \\
  & = & \rho_3 c {\mbf \sigma} \cdot \overline{\mbf p} 
    + \rho_3 \left( c {\mbf \sigma} \cdot  
    \overline{\mbf p} \: \frac{\phi}{c^2}  
    + \frac{\phi}{c^2} \: c {\mbf \sigma} \cdot \overline{\mbf p}
    \right) + \frac{1}{c^2} A + \rho_1 mc^2 \frac{\phi}{c^2} \nonumber \\
  & & + \: i \rho_2 \frac{1}{2} mc^2 \left( \frac{c {\mbf \sigma} 
    \cdot \overline{\mbf p}}{c^2 \overline{p}^2} \frac{A}{c^2}
    - \frac{A}{c^2} \frac{c {\mbf \sigma} \cdot \overline{\mbf p}}
    {c^2 \overline{p}^2} \right) + \rho_3 \frac{1}{2} m^2 c^4
    \frac{c {\mbf \sigma} \cdot \overline{\mbf p}}
    {c^2 \overline{p}^2} \nonumber \\
  & & - \: \rho_3 \frac{1}{8} m^2 c^4 \left[ \frac{1}
    {c^2 \overline{p}^2} \frac{\phi}{c^2} c {\mbf \sigma} \cdot
    \overline{\mbf p} + c {\mbf \sigma} \cdot \overline{\mbf p}
    \frac{\phi}{c^2} \frac{1}{c^2 \overline{p}^2} 
    - \left( \frac{c {\mbf \sigma} \cdot \overline{\mbf p}}
    {c^2 \overline{p}^2} \frac{\phi}{c^2} + \frac{\phi}{c^2}
    \frac{c {\mbf \sigma} \cdot \overline{\mbf p}}
    {c^2 \overline{p}^2} \right) \right] \nonumber \\
  & & - \: \frac{1}{8} m^2 c^4 \left( \frac{A}{c^2} 
    \frac{1}{c^2 \overline{p}^2} 
    - 2 \frac{c {\mbf \sigma} \cdot \overline{\mbf p}}
    {c^2 \overline{p}^2} \frac{A}{c^2} \frac{c {\mbf \sigma} 
    \cdot \overline{\mbf p}}
    {c^2 \overline{p}^2} + \frac{1}{c^2 \overline{p}^2} 
    \frac{A}{c^2} \right) .
\end{eqnarray}
Finally, we use the two unitary operators $U_{3} = e^{iS_3}$ and 
$U_{4} = e^{iS_4}$, where $S_{3}$ and $S_{4}$ satisfy
the relations
\begin{eqnarray}
  i \left[ S_{3} , \rho_3 c {\mbf \sigma} \cdot \overline{\mbf p}
    \right] & = & - \: \rho_1 m c^2 \frac{\phi}{c^2} , \\
  i \left[ S_{4} , \rho_3 c {\mbf \sigma} \cdot \overline{\mbf p}
    \right] & = & - \: i \rho_2 \frac{1}{2} m c^2 \left( 
    \frac{c {\mbf \sigma} \cdot \overline{\mbf p}}
    {c^2 \overline{p}^2} \frac{A}{c^2} - \frac{A}{c^2} 
    \frac{c {\mbf \sigma} \cdot 
    \overline{\mbf p}}{c^2 \overline{p}^2} \right) .
\end{eqnarray}
We here assume the existence of these unitary operators, which
make the remaining odd terms vanish. (We need not find the 
concrete forms of these unitary operators, because extra terms
arising from these unitary transformations are higher order terms.)
Using these unitary operators, we obtain the transformed Hamiltonian
$UH'U^{\dag}$ which is even up to the order of our interest,
\begin{eqnarray}
  U H' U^{\dag}
  & = & \rho_3 c {\mbf \sigma} \cdot \overline{\mbf p} 
    + \rho_3 \left( c {\mbf \sigma} \cdot  
    \overline{\mbf p} \: \frac{\phi}{c^2}  
    + \frac{\phi}{c^2} \: c {\mbf \sigma} \cdot \overline{\mbf p}
    \right) + \frac{1}{c^2} A  \nonumber \\
  & & \quad + \: \rho_3 \frac{1}{2} m^2 c^4
    \frac{c {\mbf \sigma} \cdot \overline{\mbf p}}
    {c^2 \overline{p}^2} \nonumber \\
  & & \quad - \: \rho_3 \frac{1}{8} m^2 c^4 \left[ \frac{1}
    {c^2 \overline{p}^2} \frac{\phi}{c^2} c {\mbf \sigma} \cdot
    \overline{\mbf p} + c {\mbf \sigma} \cdot \overline{\mbf p}
    \frac{\phi}{c^2} \frac{1}{c^2 \overline{p}^2} 
    - \left( \frac{c {\mbf \sigma} \cdot \overline{\mbf p}}
    {c^2 \overline{p}^2} \frac{\phi}{c^2} + \frac{\phi}{c^2}
    \frac{c {\mbf \sigma} \cdot \overline{\mbf p}}
    {c^2 \overline{p}^2} \right) \right] \nonumber \\
  & & \quad - \: \frac{1}{8} m^2 c^4 \left( \frac{A}{c^2} 
    \frac{1}{c^2 \overline{p}^2} 
    - 2 \frac{c {\mbf \sigma} \cdot \overline{\mbf p}}
    {c^2 \overline{p}^2} \frac{A}{c^2} \frac{c {\mbf \sigma} 
    \cdot \overline{\mbf p}}
    {c^2 \overline{p}^2} + \frac{1}{c^2 \overline{p}^2} 
    \frac{A}{c^2} \right) ,
\end{eqnarray}
where $U$ is given by $U = U_{4} U_{3} U_{2} U_{1}$.  From this, we
can derive Eq.~(\ref{ultra-rel-h1}) by a simple calculation.

\end{document}